# A New Formula of Averaging Physical Quantities

## (Application to calculation of the average radius of tapering tube and average flow velocity in the tube)

I. A. Stepanow[*]

School of Chemistry,University of Nottingham,University Park

Nottingham, NG7 2RD, UK

**Abstract**

The traditional method of finding the average value of a physical quantity often gives wrong results. Another formula of averaging is derived which gives correct results. It is applied to calculation of the average radius of tapering tube and the average flow velocity in the tube. The new formula is applicable to many other processes.

---

[*] *Email address*: istepanov@gmx.de.

## 1. Introduction

In some problems of technics and biology flow in tapering tubes is considered, for example, flow of blood in vessels. Sometimes it is useful to find the average radius of the tapering tube or, in general, that of a tube with variable radius. Consider flow of liquid in a tube which has the form of truncated cone. Its left side has radius $R_1$, its right side has radius $R_2$ and its length is L. It is necessary to find the average radius of the tube. There is the traditional equation of finding the average value:

$$\overline{f(x)} = 1/(x_2 - x_1) \int_{x_1}^{x_2} f(x)dx . \qquad (1)$$

By this equation, the average radius $\overline{R}$ is equal to $(R_1 + R_2)/2$. If $R_2 = 0$ then the flow through the tube is impossible, but Eq. (1) gives $\overline{R} = R_1/2$. The flow through the tube is proportional to the square of the average radius, therefore, according to Eq. (1), flow in the tube must exist. It is a contradiction and one sees that Eq. (1) is not applicable in this case. In this paper another equation of averaging is proposed which gives the true results.

## 2. Theory

Earlier the author worked in physics of friction. There was the following problem. It is necessary to find the average width of a clearance with roughness when lubrication flows through this clearance. The size of roughness is comparable to the width of the clearance.

Let us consider a two-dimensional clearance. Roughness is present on its lower and upper sides. The width of the clearance is denoted by h(x). The x axis is directed along the clearance. One can write

$$h(x) = h_0(x) + \varepsilon_1(x) + \varepsilon_2(x) = h_0(x) + \varepsilon(x) \qquad (2)$$

where $h_0(x)$ is the width of the clearance without roughness, $\varepsilon_1(x)$ and $\varepsilon_2(x)$ are random functions which describe the roughness on the lower and upper sides, respectively, and $\varepsilon(x) = \varepsilon_1(x) + \varepsilon_2(x)$. It was assumed that $<\varepsilon(x)> = 0$ and $<h(x)> = <h_0(x)>$ (Usov 1983, 1984, 1986; Galahov and Usov, 1990). Averaging in (Usov 1983, 1984, 1986; Galahov and Usov 1990) was done by Eq. (1). According to Eq. (1), $<h(x)>$ always equals $<h_0(x)>$ and is independent on the size of roughness. However, if $\varepsilon(x)$ equals $h_0(x)$ then liquid can not flow through the clearance. It is necessary to find a new method of averaging.

The author solved this problem like this. Consider the model problem. A ship goes along the river from A to B and back. Its speed in calm water is $v_0$ and the speed of the flow is $\Delta$. The distance from A to B is L. It is necessary to find the average speed of the ship during the travel from A to B and back. Eq. (1) gives the average speed equal to $v_0$ but this solution is wrong:

$$<v> = 1/(2L) \int_0^{2L} v(x)dx = 1/(2L)\left(\int_0^L (v_0 + \Delta)dx + \int_L^{2L} (v_0 - \Delta)dx\right) = v_0. \qquad (3)$$

The correct solution is given by the equation

$$<v> = 2L/(L/(v_0 + \Delta) + L/(v_0 - \Delta)) = v_0 - \Delta^2/v_0. \qquad (4)$$

Let us find the width of the clearance with roughness analogously. Let us suppose that the average roughness height is $\Delta/2$. Then the sum of the average roughness heights of the lower and upper sides is $\Delta$, and

$$\langle h(x)\rangle = \langle h_0(x) + \varepsilon(x)\rangle = L/(L/(2(h_0(x) - \Delta)) + L/(2(h_0(x) + \Delta))) =$$

$$h_0(x) - \Delta^2/h_0(x). \tag{5}$$

Here it is taken into account that a half of the clearance length is occupied by the peaks and the other half is occupied by the cavities. One sees that if $\Delta$ changes from $0.1h_0(x)$ to $h_0(x)$ then $\langle h(x)\rangle$ changes from $0.99h_0(x)$ to 0.

The general equation of finding the average speed of the ship will be

$$\langle v\rangle = 2L/\int_0^{2L} dx/v(x) \tag{6}$$

and the general equation of the new type of averaging will be

$$\langle f(x)\rangle = (x_2 - x_1)/\int_{x_1}^{x_2} dx/f(x). \tag{7}$$

Let us use Eq. (7) for calculation of the average radius of a tapering tube. The dependence of the radius on the distance along the axis of the tube is $R(x) = (R_2 - R_1)/L \cdot x + R_1$ and $x_2 - x_1 = L$ hence

$$\langle R\rangle = (R_2 - R_1)/\ln R_2/R_1. \tag{8}$$

It is necessary to check the validity of Eq. (7). One can calculate the average velocity of liquid in the tube by Eqs. (1) and (7). According to the definition, the average velocity is the velocity whose displacement for the time t is equal to the displacement for the variable velocity for the same time:

$$<v>t = L. \tag{9}$$

Using the continuity equation

$$R_1^2 v_1 = R_2^2 v_2 \tag{10}$$

($v_1$ and $v_2$ are fluid velocities at the inlet and at the outlet of the tube, respectively) one can calculate $\bar{v}$ by Eq. (1) and $<v>$ by Eq. (7):

$$\bar{v} = R_1/R_2 v_1 \tag{11}$$

and

$$<v> = 3 R_1^2 v_1 (R_2 - R_1)/(R_2^3 - R_1^3). \tag{12}$$

The time t can be found from the equation

$$t = \int_0^L dx/v(x), \tag{13}$$

namely

$$t = L(R_2^3 - R_1^3)/3R_1^2 v_1(R_2 - R_1). \tag{14}$$

It is clear that $\bar{v}t \neq L$ but $<v>t = L$.

The Bernoulli's equation

$$P_1 + 0.5\rho v_1^2 = P_2 + 0.5\rho v_2^2. \tag{15}$$

Using it one can calculate $\bar{v}$ from Eq. (1) and $<v>$ from Eq. (7). One can easy show that $\bar{v}t \neq L$. However,

$$<v> = L/\int_0^L dx/v(x) \tag{16}$$

and using Eq. (13) it is clear that $<v>t \equiv L$.

Let us solve the following problem. It is necessary to find the volume of liquid flow through tapering tube. For a tube with constant radius, the Poiseuille's equation gives

$$Q = \Pi R^4 (P_1 - P_2)/(8\mu L) \tag{17}$$

where $\mu$ is the dynamic viscosity of the liquid. One can try to calculate the volume flow through tapering tube by the following formula:

$$Q = \Pi <R^4>(P_1 - P_2)/(8\mu L) \tag{18}$$

where $<R^4>$ is the average value of $R^4$. One can show that averaging using Eq. (1) leads to wrong result. Averaging using Eq. (7) gives

$$<R^4> = L/\int_0^L dx/R^4 = 3(R_1 - R_2)/(1/R_2^3 - 1/R_1^3). \quad (19)$$

Introducing it into Eq. (18), one gets

$$Q = 3\Pi(R_1 - R_2)(P_1 - P_2)/(8\mu L(1/R_2^3 - 1/R_1^3)). \quad (20)$$

One can find a few exact solutions to this problem. One of them is to treat a thin tapering liquid shell of thickness dr, inner radii $r_1$ at the left edge and $r_2$ at the right edge, and length L; its axis is coincident with the axis of the tube, then to consider pressure and viscous forces acting on it. The simplest method is: For infinitely small element of tube with the length dx, the equation (17) can be used if $(P_1 - P_2)/L$ is replaced by - dP/dx:

$$Q = -\Pi R^4 dP/(8\mu dx). \quad (21)$$

Then

$$-dP = 8\mu Q/(\Pi R^4)dx \quad (22)$$

and

$$P_1 - P_2 = 8\mu Q/\Pi \int_0^L dx/R^4 = 8\mu Q/(3\Pi)L/(R_1 - R_2)(1/R_2^3 - 1/R_1^3). \qquad (23)$$

From Eq. (23), equation (20) follows. One sees that the exact method is more tedious.

One can try to use this method for calculation of the average radius of elastic tubes with varying diameter (for example, blood vessels) and for simple calculation of the volume of liquid flow through such tubes.

### 3. Conclusions

One can make the following conclusion. The traditional equation of averaging (1) gives wrong results for calculation of average values of physical quantities in tubes with flow of liquid. The proposed equation (7) gives correct results. The reason is that Eq. (1) does not take into account interaction of liquid with the walls of the tube.